\documentclass[draftcls,12pt,onecolumn]{IEEEtran}
 \normalsize

\usepackage[cmex10]{amsmath}
\usepackage{cases, cite, graphicx, amsfonts, amssymb, amsthm, algorithm, algorithmic, psfrag, pstool, verbatim, relsize, color}

\newtheorem{lemma}{Lemma}

\newtheorem{theor}{Theorem}

\newcommand{\argmax}{\operatornamewithlimits{argmax}}
\newcommand{\Prm}{P_{\mathrm{r}}}
\newcommand{\Rc}{R_{\mathrm{c}}}
\newcommand{\Rs}{R_{\mathrm{s}}}
\newcommand{\Xr}{X_{\mathrm{r}}}

\begin{document}

\title{Maximizing Data Rate for Multiway Relay Channels with Pairwise Transmission Strategy}

\author{Reza~Rafie~Borujeny,~\IEEEmembership{Student Member,~IEEE, }%
        Moslem~Noori,~\IEEEmembership{Member,~IEEE,}%
        
        and~Masoud~Ardakani,~\IEEEmembership{Senior~Member,~IEEE}%
        \thanks{The authors are with the Department of Electrical and Computer Engineering, University of Alberta, Edmonton, AB, Canada T6G 2V4
        (e-mail: \{reza.rafie, moslem, ardakani\}@ualberta.ca).}%
        \thanks{A portion of this work was presented in ISIT 2014.}%
}

\maketitle

\vspace{-1.5cm}
\begin{abstract}
In a multiway relay channel (MWRC), pairwise transmission strategy can be used to reduce the computational complexity at the relay and the users without sacrificing the data rate, significantly. The performance of such pairwise strategies, however, is affected by the way that the users are paired to transmit. In this paper, we study the effect of pairing on the common rate and sum rate of an MWRC with functional-decode-forward (FDF) relaying strategy where users experience asymmetric channel conditions. To this end, we first develop a graphical model for an MWRC with pairwise transmission strategy. Using this model, we then find the maximum achievable common rate and sum rate as well as the user pairings that achieve these rates. This marks the ultimate performance of FDF relaying in an MWRC setup. Further, we show that the rate enhancement achieved through the optimal user pairing becomes less pronounced at higher SNRs. Using computer simulations, the performance of the optimal pairing is compared with those of other proposed pairings in the literature.
\end{abstract}


\IEEEpeerreviewmaketitle
\makeatletter

\section{Introduction}\IEEEPARstart{A}{ multiway} relay channel (MWRC) \cite{Gunduz2009} is an extension of a two-way relay channel \cite{TWRC1,Wilson,TWRC3,TWRC4,Popovski} in which $N \geq2$ users intend to share their data. This could be partial data sharing, where each user shares its data with not all but a subset of other users, or full data sharing when each user share its data with all other users. It is common to assume that no direct link is present between the users and a relay assists them to communicate their messages. Conference calls, file sharing, and multi-player gaming \cite{Moslem, MIMO} are potential applications of MWRCs. Different from conventional cooperative schemes, each user serves as both data source and data destination in an MWRC. This means that we have simultaneous data flows in different directions necessitating the design of customized transmission strategies for MWRCs. 

\emph{Pairwise transmission}  \cite{Gunduz2009, Moslem, Cadambe, Ong_Binary,Ong}, also known as \emph{pairwise network coding}, is one of the main transmission strategies proposed for MWRCs. To accomplish full data exchange based on pairwise transmission, a set of pairs, representing the users' transmission schedule, is defined. Every two users within a pair simultaneously send their data to the relay in an uplink phase. In \emph{functional-decode-forward} (FDF) \cite{Ong_Binary}, the relay directly decodes a function of the two received messages. Following the uplink phase, the relay broadcasts the function of the two users' messages to all users \cite{Gunduz2009}  in a downlink phase. The uplink and downlink transmissions continue until each user is capable of decoding all other users' messages. It is worth mentioning that pairwise transmission has a low decoding complexity, while offering interesting capacity-achieving properties in various MWRC setups \cite{Ong_Binary,Ong}. For instance, it has been shown that pairwise transmission along with rate splitting and joint source-channel decoding achieves the capacity region of MWRC over finite fields \cite{Ong}. 

In an MWRC with pairwise transmission, the way that users are paired for transmission, referred to as \emph{users' pairing}, directly affects the achievable data rates of the system \cite{Moslem}. That said, the effect of pairing on the common rate (the rate that any user can reliably transmit its data with this rate to all other users) of an MWRC has been the subject of various studies. Considering different constraints on the relay transmit power, authors in \cite{equalrate} have shown that their pairing strategy maximizes the common rate for an MWRC with FDF relaying where each user's transmitted signal can depend on both its own message as well as its previously received signals. In \cite{Moslem}, authors have found the optimal pairing to maximize the achievable common rate of the users for an MWRC with asymmetric Gaussian channels under the assumption that each user transmits in at most two uplink phases. In \cite{Tao}, an opportunistic approach for finding the pairing in a pairwise transmission for MWRC with compute-and-forward relaying has been proposed. Further, \cite{Parastoo} has considered a pairing in which the user with the highest SNR is paired with all other users and the common rate and sum rate of the system have been investigated for various channel configurations.

In this work, we seek optimal pairings to maximize the common rate and sum rate for FDF relaying. Thus, the ultimate rate performance achieved by FDF is determined. To this end, we first introduce a graph-based modeling for the data transmission in a pairwise MWRC. Using this model, we then find the necessary and sufficient conditions for a pairing to be feasible, i.e., each user is able to retrieve the data of all other users. Using this condition, we then discuss that there exist $N^{N-2}$ distinct feasible pairings in the system. Thus, finding the optimal pairing through brute-force search becomes extremely expensive as $N$ increases. That said, it is desired to analytically find the optimal pairings. To address this, we use the developed graph-based model to analytically find  common rate and sum rate maximizing pairings. 

The rest of the paper is organized as follows: In Section \ref{system}, we describe the system model. In Section \ref{problem}, we introduce a graphical representation of the transmission pairing. Then, we describe the sum rate and the common rate maximization problems that we want to solve. Our proposed graphical model is used to find the solutions to these problems in Section \ref{solution}. We compare the performance of our proposed pairings with those of other transmission strategies in the literature via simulations in Section \ref{simulations}. Finally, Section \ref{conclusion} concludes the paper\footnote{This work was partially presented at the IEEE International Symposium on Information Theory, Honolulu, HI, USA, 2014 \cite{Reza}. In \cite{Reza}, the results for optimal pairing has been presented without proofs. The current work also extends the sum rate results of \cite{Reza} to cases that some users are listeners only and do not participate in data sharing.}.

\section{System Model} \label{system} 
\begin{figure}[!t]
\psfrag{U1}{$U_1$}
\psfrag{U2}{$U_2$}
\psfrag{UN}{$U_N$}
\psfrag{Relay}{$\mathcal{R}$}
\psfrag{C1r}{$C_{1\mathcal{R}}$}
\psfrag{Cr1}{$C_{\mathcal{R}1}$}
\psfrag{C2r}{$C_{2\mathcal{R}}$}
\psfrag{Cr2}{$C_{\mathcal{R}2}$}
\psfrag{CrN}{$C_{\mathcal{R}N}$}
\psfrag{CNr}{$C_{N\mathcal{R}}$}
\psfrag{...}{$\dots$}
\centering
\includegraphics[scale = 0.3]{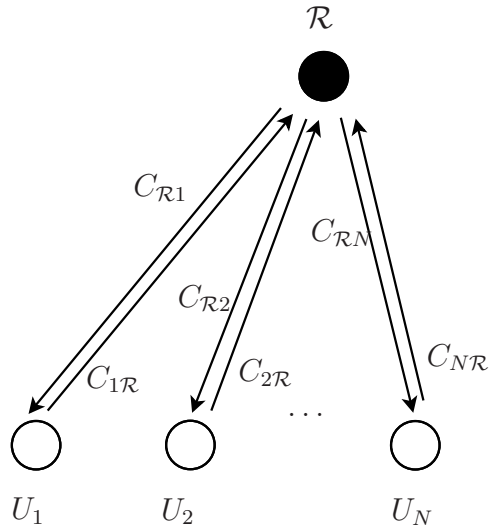}
\caption{An MWRC with $N$ users.}
\label{fig_model}
\end{figure}

We consider an MWRC in which $N$ \textcolor{black}{single-antenna} users, namely $U_1, U_2, \dots, U_N$, perform full data exchange meaning that each user wants to send/receive data to/from all other users. It is assumed that users cannot communicate directly, thus, a \textcolor{black}{single-antenna} relay $\mathcal{R}$ assists them to share their data (Fig. \ref{fig_model}). \textcolor{black}{Here, all $U_i$'s transmit their (encoded) message $X_i$'s with equal power $P$ and $\Prm$ is  the transmit power of the relay. The channel from $U_i$ to $\mathcal{R}$ is assumed to be reciprocal and slow-fading so that the channel gain $h_i$ remains unchanged during the data exchange between the users. Further, additive white Gaussian noise (AWGN) with variance $\sigma^2$ is assumed at the relay and users.}

The uplink signal to noise ratio (SNR) for user $U_i$, namely $\gamma_i$, is defined as 
$\gamma_i \triangleq \frac{P \vert h_ i \vert^2}{\sigma^2}$. \textcolor{black}{Without loss of generality, we assume that $\vert h_1 \vert \leq \vert h_2 \vert \leq \ldots \vert h_N \vert$, and thus\footnote{\textcolor{black}{Note that for any given MWRC with $N$ users, we can always label the user with the worst SNR (poorest channel) as $U_1$, the second worst user as $U_2$, and so on. Thus, for any MWRC, we can always have  $\vert h_1 \vert \leq \vert h_2 \vert \leq \ldots \vert h_N \vert$ via a simple relabeling of the users.}}} 
\begin{equation} \label{assum}
\gamma_N \geq \gamma_{N-1} \geq \dots \geq \gamma_1 > 0.
\end{equation}
Similarly, a user downlink SNR is defined as $\Gamma_i \triangleq \frac{\Prm\vert h_i \vert^2}{\sigma^2}$. We denote the minimum of downlink SNRs by $\Gamma_{\mathrm{d}} = \min_i \{\Gamma_i\}$. \textcolor{black}{Note that since $\min_i \vert h_i \vert = \vert h_1 \vert$, $\Gamma_{\mathrm{d}} = \Gamma_{1}$.}

\subsection{Pairwise Transmission Strategy}
\textcolor{black}{In a pairwise transmission scheme, users are grouped into $M = N - 1$ pairs\footnote{\textcolor{black}{Note that here, any of the users wants to multicast its data to all other users. Here, the term pairwise refers to the transmission strategy in a MWRC and should not be confused with multi-user or multi-pair two way relaying \cite{tian2014degrees, wang2014mimo, liu2015generalized, kim2010capacity} where several unicast message exchanges happen between the users.}}.} These pairs are not necessarily disjoint meaning that a specific user can appear in more than one pair. However, all messages sent by a user in a round of communication are identical. This is necessary to ensure successful decoding at the users as discussed later. Such a set of pairs is called a \textit{pairing} of the users and is denoted by an \textcolor{black}{$M$-tuple} $O = (\{u_{11},  u_{12}\}, \dots, \{u_{M1}, u_{M2}\})$ where $u_{\ell1}$ and $u_{\ell2} \in \{U_1, U_2, \dots, U_N\}$ for any $\ell \in \{1, \ldots, M \}$.

\textcolor{black}{Now that we have defined $O$, we explain how the users share their data over a data exchange round, i.e. $M$ uplink phases and $M$ downlink phases}. In the $\ell$th uplink phase, $\ell \in \{1,\ldots,M \}$, the users associated with the $\ell$th pair of the pairing simultaneously send their data to the relay. \textcolor{black}{Let us call the users in this pair by $U_i$ and $U_j$ that apply a coding scheme defined over a field $\mathbb{F}$ to produce their coded messages (vectors) $X_i$ and $X_j$ respectively. The uplink phase is then followed by a downlink phase where the relay broadcasts the sum of the two messages received in the last uplink phase to the users. That is, the relay broadcasts $X_i \oplus X_j$ in the downlink phase where $\oplus$ refers to the element-wise summation of $X_i$ and $X_j$ over $\mathbb{F}$.} The discussion on how the relay forms its transmit messages is discussed later.

These pairwise transmissions continue until the last pair of the pairing. After the last downlink transmission, each user has a set of $M$ equations where each equation is a linear combination of two users' messages.  By having the knowledge of self message, $U_i$ attempts to solve its received set of equations successfully. Note that the transmit power of the users is fixed for during each uplink phase of an each exchange round. For more details please see \cite{Ong_Binary}. If the system of $M$ equations at each user is solvable, we say that the corresponding pairing is \textit{feasible}. The notion of feasibility ensures that each user can find all other messages. 

%


A clear advantage of a pairwise relaying over joint multi-user decoding (e.g. full decode and forward), is its lower complexity. While the decoding complexity of joint decoding grows exponentially with the number of users \cite{Verdu1, Verdu2}, in a pairwise system the complexity \textcolor{black} {grows linearly} with $N$ (or as constant when normalized by the number of users). This is because each user decodes the message of other users one by one. Lower decoding complexity also benefits the relay where it only needs to deal with the message of only two users at a time regardless of $N$. One should note that for the pairwise system to work, each user has to know the pairing schedule. This can be broadcast to the users by the relay.  

\subsection{Achievable Data Rates}
Users employ channel codes to protect their data against the noise. For the assumed MWRC with $N$ users, a $(2^{nR_1}, 2^{nR_2}, \ldots, 2^{nR_N}, n)$ code consists of the following four components:

\begin{itemize}
\item Users' messages: They are represented by $N$ sets of integers $\mathcal{W}_i = \{1,2,\ldots, 2^{nR_i}\}$ for $i = 1,\ldots, N$. Each set $\mathcal{W}_i$ represents the $U_i$'s original messages. 
\item Users' encoding functions: An encoding function $f_i(\cdot)$ is assigned to $U_i$ that takes a $W_i \in \mathcal{W}_i$ and forms $X_i$ as $X_i = f_i(W_i)$. 
\item Relay's encoding function: The relay's transmit message at the $\ell$th downlink phase, $X_{\mathrm{r}}^{\ell}$, is formed by the encoding function $f_{\mathrm{r}(\cdot)}$ as $X_{\mathrm{r}}^{\ell} = f_{\mathrm{r}}(Y_{\mathrm{r}}^{\ell})$ where $Y_{\mathrm{r}}^{\ell}$ is the relay's received signal from the $\ell$th uplink phase. 
\item Users' decoding functions: The decoding function at $U_i$ uses the received signals from all downlink phases at $U_i$ as well as the knowledge of self message to decode the data of all other users. In other words, 
\begin{equation}
(\hat{W}_1,\ldots, \hat{W}_N) = g_i(Y_i^1,\ldots, Y_i^{M}, W_i)
\end{equation}
where $Y_i^{\ell}$ is the received signal at $U_i$ in the $\ell$th downlink phase and $\hat{W}_j$ is the estimate of $W_j$.
\end{itemize}

For the aforementioned code, the average probability of error is 
\begin{equation}
P_{e}^{n}={\rm Pr}\bigcup_{i = \{1,\ldots,N\}} \left\{g_i(Y_i^1,\ldots, Y_i^{M}, W_i) \ne(W_{1},\ldots, W_{N})\right\}.
\label{eq:}
\end{equation}
Now, a rate tuple $(R_1, R_2, \dots, R_N)$ is said to be achievable if there exists a code where $P_{e}^{n} \rightarrow 0$ as $n$ goes to infinity. In other words, any $U_i$ can reliably (with arbitrarily small probability of error) transmit its data to all other users with rate $R_i$ after each round's $M$ uplink and downlink phases (i.e., a complete round of full data exchange). Knowing that $(R_1, R_2, \dots, R_N)$ is achievable, all users can reliably share their data with a common rate $\Rc$ where
\begin{equation}
\Rc \triangleq \min_i {R_i}.
\end{equation} 
Also, the sum rate $\Rs$ is defined as 
\begin{equation}
\Rs \triangleq \sum_{i=1}^{N}R_i.
\end{equation}

\subsection{FDF Relaying}
\color{black}
Assume that at the $\ell$th uplink phase $u_{\ell1} = U_i$ and $u_{\ell2}=U_j$ are paired and transmit their data to the relay. As a result, the received singal at the relay is
\begin{equation}
Y_{\mathrm{r},\ell} = h_i X_i + h_j X_j + Z_{\mathrm{r}}
\end{equation}
where $Z_{\mathrm{r}}$ is the AWGN at the relay. After receiving $Y_{\mathrm{r},\ell}$, the relay forms its message $X_{\mathrm{r},l} = X_{i} \oplus X_{j}$ to be transmitted to all users in the $\ell$th downlink phase. \color{black}
To form $X_{i} \oplus X_{j}$, the relay uses nested lattice codes \cite{Wilson, Nam2010} at the users for a more efficient decoding at the relay. The basic notion of this technique is that by using lattice codes,  the relay is capable of directly decoding the summation of the received messages. It means that it decodes $X_{i} \oplus X_{j}$ rather than separately decoding $X_i$ and $X_j$. For more details on FDF, see \cite{Nam2010, equalrate}. \color{black}
After the relay's transmission in the $\ell$th downlink phase, the received signal at an arbitrary user $k$ is
\begin{equation}
Y_{k,\ell} = h_k X_{\mathrm{r},l} + Z_k
\end{equation}
where $Z_k$ is the AWGN at $U_k$.\color{black}

Having the above transmission model and using the results in \cite{Nam2010, Moslem}, The achievable rates of $U_i$ and $U_j$, denoted by $R_i$ and $R_j$ respectively, are limited by the following achievable upper bounds
\begin{align}\label{FDF_rate}
R_i \leq \max\left\{0, \frac{1}{2M}\log_2\left(\frac{\gamma_i}{\gamma_i + \gamma_j}+\gamma_i\right)\right\}, \\ \label{FDF_rate_2}
R_j \leq \max\left\{0, \frac{1}{2M}\log_2\left(\frac{\gamma_j}{\gamma_i + \gamma_j}+\gamma_j\right)\right\}.
\end{align} 

In addition, the transmission rates of the users may be limited by the downlink phase. More specifically, the transmit rate of any $U_i$ is bounded as follows
\begin{equation}\label{dlub}
R_i \leq \frac{1}{2M}\log_2\left(1+ \Gamma_{\mathrm{d}}\right).
\end{equation}
\textcolor{black}{Following the above, the overall upper bound on $R_i$ is now found by taking the minimum of all uplink upper bounds on $R_i$ and its downlink upper bound. }

\textcolor{black}{As seen from (\ref{FDF_rate}), $R_i$, and as a consequence the common rate and sum rate of the system, are functions of $\frac{\gamma_i}{\gamma_i + \gamma_j}$. On the other hand, since the choice of the users' pairing affects $\frac{\gamma_i}{\gamma_i + \gamma_j}$, both common rate and sum rate are affected by the users' pairing. This motivates us to seek pairings that maximize $\Rc$ and $\Rs$. In the following section, we explain the problem of finding such pairings in more detail. }

\section{Problem Definition} \label{problem} In this section, we first introduce the concept of client graph which is a graphical description of users' pairing. Then, we define the problems that we study in this work. \textcolor{black}{As we assume that all channel gains remain unchanged during each round of full data exchange, it does not matter which of the available $M$ uplink phases are allocated to a pair. That is, while the way that the pairs are formed affects the common and sum rate, allocation of the uplink phases to the pairs is irrelevant to the data rate. Thus, instead of using an $M$-tuple $O = (\{u_{11},  u_{12}\}, \dots, \{u_{M1}, u_{M2}\})$ to denote a pairing, we use a set representation as $O = \{\{u_{11},  u_{12}\}, \dots, \{u_{M1}, u_{M2}\}\}$ in the rest of the paper.}

\subsection{Client Graph} \label{graph} 
An undirected graph $G$ is an ordered pair $G = (V, E)$ comprising a set $V = \{v_1, v_2, \dots, v_K\}$ of vertices together with a set $E$ of edges. For simplicity, if $\{v_i, v_j\}\in E$, we say $v_iv_j\in E$. If $v_iv_j\in E$, we say $v_j$ is \textit{adjacent} to $v_i$. The set of adjacent vertices of $v_i$, denoted by $A_i^G$, is called the set of neighbors of $v_i$. Also, the \textit{degree} of node $v_i$ is $deg(v_i) = |A_i^G|$. The adjacency matrix of $G$, denoted by $\mathcal{A} = (a_{ij})$, is a $K\times K$ matrix in which $a_{ij} = 1$ if and only if (iff) $v_iv_j\in E$; otherwise $a_{ij} =0$. A \textit{path} in $G$ is a sequence of consecutive edges that connects a sequence of vertices. $G$ is called connected if there is at least one path between every pair of its vertices. A non-empty path with the same endpoints is called a \textit{cycle}.

For a given pairwise pairing $O$, we define a \textit{client graph} $G_O(V, E)$ where $V = \{v_1, v_2, \dots, v_N\}$ is the set of vertices. There is a vertex $v_i$ in $V$ corresponding to each user $U_i$. There exists an edge $e = v_iv_j \in E$ iff $\{U_i, U_j\} \in O$. Note that there is a one-to-one correspondence between all possible client graphs and all possible pairwise pairings. \textcolor{black}{As an example, Figure~\ref{CRFDFgraph} depicts the client graph associated with pairing $O_{\Rc} = \{\{U_1,U_2\},\{U_2,U_3\}, \{U_3,U_4\},\dots ,\{U_{N-1},U_N\}\}$.}


\begin{theor}\label{th1}
A pairing with $M=N-1$ pairs is feasible iff the corresponding client graph is a tree\footnote{\textcolor{black}{In graph theory, a tree refers to a graph that does not have any cycle.}}.
\end{theor}

\begin{IEEEproof}
For the forward direction, note that if the client graph is not a tree, then it has $k>1$ components. That is, the corresponding system of linear equations consists of $k$ uncoupled systems of equations. This contradicts the feasibility of the pairing. For the backward direction, we use the fact that if the client graph is a tree, then there is exactly one path $P_{i,j}$ between any pair of nodes $v_i$ and $v_j$. Assume that $P_{i,j} = \{v_{i}v_{i_1}, v_{i_1}v_{i_2}, \dots , v_{i_{n-1}}v_{j}\}$. The equations corresponding to the edges in this path are:
\begin{IEEEeqnarray}{c}\label{syseq}
X_{i} \oplus X_{i_1} = \Xr^{m_1}\IEEEnonumber\\
X_{i_1} \oplus X_{i_2} = \Xr^{m_2}\\\IEEEnonumber
\vdots \\\IEEEnonumber
X_{i_{n-1}} \oplus X_{j} = \Xr^{m_n}\IEEEnonumber
\end{IEEEeqnarray}
in which $\Xr^{m_k}$ represents the relay message at the corresponding downlink.
Manipulating this system of equations, we wind up with
\begin{equation} \label{Eq Tree proof}
X_{i} \oplus (-1)^{n-1}X_{j} = \bigoplus_{k = 1}^{n}{(-1)^{k-1} \Xr^{m_k}}.
\end{equation}
Knowing its own data, $U_i$ can decode $X_j$ for all $j\neq i$ according to (\ref{Eq Tree proof}). Thus, if the client graph is a tree, the corresponding pairing is feasible.
\end{IEEEproof}

In the following, we use the terms \emph{client tree} and client graph, interchangeably. Further, we denote the maximum achievable common rate and sum rate for a client graph $G_O$ by $\Rc(G_O)$ and $\Rs(G_O)$, respectively.

\subsection{Common Rate and Sum Rate Maximization}
In this paper, we focus on two problems related to the data rate performance of the considered MWRC. First, we consider \textit{common rate maximization problem} meaning that we are interested in finding a pairing that maximizes $\Rc$. More formally, if we denote the set of all feasible pairings with $\mathcal{O}$, the optimal pairing is defined as
\begin{equation}\label{GenCR}
O_{\Rc} = \argmax_{O \in \mathcal{O}}{\Rc(G_O)}
\end{equation}

The second problem is \emph{sum rate maximization problem} where we want to find the pairing that maximizes the sum rate of the considered MWRC. To address this problem, we assume that all users want to participate in each round of full data exchange and transmit data with a non-zero rate. This can be interpreted as a level of fairness \cite{Parastoo} between all users such that we do not push any of them to stop transmitting. However, we will discuss how this can be extended to a more general setup where some users may stay silent in an exchange round. Furthermore, in the sum rate maximization problem, it is assumed that achievable rates are not limited by the downlink phase and we only focus on the effect of the pairing on the achievable rates of uplink. To be more specific, we are interested to find an optimal pairing such that 
\begin{equation}\label{GenSR}
O_{\Rs}=\operatorname*{arg\,max}_{O \in \mathcal{O}}{\Rs(G_O)}.
\end{equation}

In order to solve a common (sum) rate maximization problem, we need to find a client graph $G_O$ with greatest $\Rc(G_O)$ ($\Rs(G_O)$) among all client trees. One way is to search over all possible client trees and find the one that maximizes $\Rc(G_O)$ ($\Rs(G_O)$). According to Cayley's formula \cite{cayley}, this approach needs searching over  $N^{N-2}$ client trees which is impractical even if the number of users is not very large. This motivates us to develop efficient solutions for finding the optimal client trees without going through such tedious searches.

\section{Problem Solution} \label{solution} In this section, we provide solutions to the common rate and sum rate maximization problems as defined by (\ref{GenCR}) and (\ref{GenSR}) for FDF relaying. We emphasize that the optimality considered in this section is limited to finding the best choice for pairing in a pairwise transmission strategy.

\begin{figure}[!t]
\psfrag{U_1}{$U_1$}
\psfrag{U_2}{$U_2$}
\psfrag{U_3}{$U_3$}
\psfrag{U_N}{$U_{N}$}
\psfrag{...}{$\dots$}
\centering
\includegraphics[scale = 0.3]{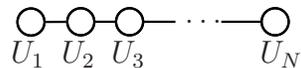}
\caption{Client tree that maximizes $\Rc(G_O)$ for a pairwise MWRC with FDF relaying.}
\label{CRFDFgraph}
\end{figure}

\subsection{Common Rate Maximization}
Considering (\ref{FDF_rate}), we find the pairing that achieves the maximum $\Rc(G_O)$ for FDF relaying.
Theorem \ref{FDFCR} gives the optimal pairing for this scenario. 
\begin{theor}\label{FDFCR}
The pairing given by
\if@twocolumn
\begin{equation}
\begin{split}
&O_{\Rc} = \\
&\{\{U_1,U_2\},\{U_2,U_3\}, \{U_3,U_4\},\dots ,\{U_{N-1},U_N\}\}
\end{split}
\end{equation}
\else
\begin{equation}
O_{\Rc} = \{\{U_1,U_2\},\{U_2,U_3\}, \{U_3,U_4\},\dots ,\{U_{N-1},U_N\}\}
\end{equation}
\fi
achieves the maximum common rate in an MWRC with FDF relaying and the maximum achievable common rate is
\if@twocolumn
\begin{equation}
\begin{split}
&\Rc(G_O) = \\
&\! \frac{1}{2(N-1)}\! \min_{i \! \in \! \{1,\ldots,N\!\}\!} \!^+  \!\!\left\{ \! \! \log_2\left( \! \gamma_i \! +\!  \frac{\gamma_i}{\gamma_i \! +\! \gamma_{i+1}} \! \right)  \!, \log_2\left(1+ \Gamma_{\mathrm{d}} \right) \! \right\}.
\end{split}
\end{equation}
\else
\begin{equation}
\Rc(G_O)  =  \frac{1}{2(N-1)}\min_{i  \in  \{1,\ldots,N\}}\!^+ \!\!\left\{  \log_2\left(  \gamma_i  +  \frac{\gamma_i}{\gamma_i  + \gamma_{i+1}}  \right)  , \log_2\left(1+ \Gamma_{\mathrm{d}} \right)  \right\}.
\end{equation}
\fi
in which $\min^+ A = \max\{0, \min A\}$.
\end{theor}
\begin{IEEEproof}
See Appendix \ref{appFDFCR}.
\end{IEEEproof}
Fig. \ref{CRFDFgraph} illustrates the client graph to achieve maximum $\Rc$ with FDF relaying.
Using the results in \cite{Moslem}, it can been shown that asymptotically, as uplink SNRs increase, the performance of random pairing achieves the performance of optimal pairing.
\begin{figure}[!t]
\psfrag{U_1}{$U_1$}
\psfrag{U_2}{$U_2$}
\psfrag{U_3}{$U_3$}
\psfrag{U_N}{$U_N$}
\psfrag{...}{$\dots$}
\centering
\includegraphics[scale = 0.3]{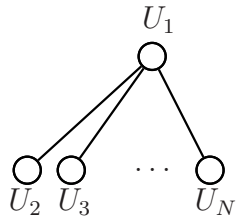}
\caption{Client tree that maximizes $\Rs(G_O)$ for a pairwise MWRC with FDF relaying subject to the weakened upper bound given by (\ref{FDF_weak}).}
\label{SRgraph}
\end{figure}

\subsection{Sum Rate Maximization}
To find a pairing with maximum sum rate, we make two assumptions. First, the downlink does not limit the rate. Second, the transmit SNRs are not too low. More specifically,  for any $i$ and $j$, $\frac{\gamma_i}{\gamma_i + \gamma_j}+\gamma_i \geq 1$. This assumption generally holds in most practical setups where the signal power is stronger than the noise power. To this end, the upper bound on achievable rate of $U_i$, when it is paired with $U_j$, is given by:
\begin{equation}\label{FDF_weak}
R_i \leq \frac{1}{2(N-1)}\log_2\left(\frac{\gamma_i}{\gamma_i + \gamma_j}+\gamma_i\right).
\end{equation}
Now, the optimal pairing is given by the following theorem:
\begin{theor}\label{FDFSR}
The pairing
\begin{equation}
O_{\Rs} = \{\{U_2,U_1\},\{U_{3},U_1\},\dots ,\{U_{N},U_1\}\}
\end{equation}
is the optimal pairing for FDF relaying subject to (\ref{FDF_weak}). Moreover, the maximum achievable sum rate for this pairing is:

\if@twocolumn
\begin{align}\label{maxSRFDF} \nonumber
\Rs(G_O) \!= \!\frac{1}{2(N\!-\!1)} \log_2 & \Bigg(\left(\gamma_1 + \frac{\gamma_1}{\gamma_1 + \gamma_N}\right) \\  
& \times \! \!  \prod_{i=2}^{N} \! \frac{\gamma_i}{\gamma_i \! +\!  \gamma_1} \! + \! \gamma_i  \! \Bigg).
\end{align}
\else
\begin{equation}\label{maxSRFDF}
\Rs(G_O) = \frac{1}{2(N-1)} \log_2  \Bigg( \left(\gamma_1 + \frac{\gamma_1}{\gamma_1 + \gamma_N}\right)
 \times   \prod_{i=2}^{N}  \left (\frac{\gamma_i}{\gamma_i  +  \gamma_1}  +  \gamma_i \right) \Bigg).
\end{equation}
\fi
\end{theor}

\begin{IEEEproof}
See Appendix \ref{appFDFSR}.
\end{IEEEproof}

\color{black}
\textbf{Remark 1:} In Theorem~\ref{FDFSR}, we assume that all the users want to transmit their data during each round of full data exchange. In the scenarios that some of the users have very poor channel conditions, one may be able to achieve higher sum rates than (\ref{maxSRFDF}) by silencing weak users and forcing them to only listen. This could potentially allow for more channel utilization by the stronger users. Even for such scenarios, the results of Theorem~\ref{FDFSR} can be effectively used to find the set of the users that should remain active and their pairing which results in the maximum sum rate of the system. We do this through running an exhaustive search where at each iteration, we force $i \in \{1,2, \ldots, N - 1\}$ users to stay silent and find the maximum sum rate. At the end of this search, we are able to determine which users should remain silent to achieve the maximum sum rate. To find the complexity of this exhaustive search, one should note that to achieve possible rate improvement by silencing $i$ users, it can be shown using (\ref{maxSRFDF}) that these $i$ users should be chosen from $S = \{U_1, U_2, \dots, U_i, U_{i+1},U_N\}$. Checking all possibilities of choosing $i$ users from $S$ has a complexity of $O(N^2)$. For each choice of $i$ users from $S$, the optimal ordering for the other $N - i$ active users is found using Theorem~\ref{FDFSR}. Note that feasibility requires having $N-i$ uplink phases in this case so that the silent users can also decode the messages. The first $N-i-1$ uplink phases can be assigned to the pairwise pairing of the $N-i$ transmitting users. In the last uplink phase, one of the users, say $U_j$, simply transmits its data to the relay. In the last downlink phase, the relay broadcasts $X_j$ to all users. It can be easily seen that this extra uplink phase required for ensuring feasibility does not change the optimal pairing. Further, since
\begin{equation}
\log_2(1 + \gamma_j) > \log_2 \left( \frac{\gamma_j}{\gamma_j + \gamma_k} + \gamma_j \right)
\end{equation}
for any user $k$, the bound on the individual uplink transmission by $U_j$ in the last phase is always larger than its uplink bound in a pairwise transmission. Thus, any of the $N - i$ active users can be selected to transmit individually in the last phase without affecting the sum rate. Here, when using (\ref{maxSRFDF}) to calculate the sum rate, the pre-log factor in (\ref{maxSRFDF}) for each subset of size $N-i$ should be $\frac{1}{2(N-i)}$ instead of  $\frac{1}{2(N-1)}$. Each iteration of exhaustive search requires quadratic time complexity in $i$. Thus, the overall complexity of an exhaustive search over all values of $i$ is $O(N^3)$, that is, polynomial in $N$.

\color{black}
\textbf{Remark 2:} From Theorem \ref{FDFSR}, one can show that the maximum achievable sum rate for the optimal pairing is
\if@twocolumn
\begin{equation}
\begin{split}
&\Rs(G_O) = \frac{1}{2(N-1)}\times\\&\log_2\left(\left(\prod_{i=1}^{N}{\gamma_i}\right)\left(\prod_{i=2}^{N}{1+ \frac{1}{\gamma_i+\gamma_1}}\right)\left(1+\frac{1}{\gamma_1+\gamma_N}\right)\right).
\end{split}
\end{equation}
\else
\begin{equation}
\Rs(G_O) = \frac{1}{2(N-1)}\times\log_2\left(\left(\prod_{i=1}^{N}{\gamma_i}\right)\left(\prod_{i=2}^{N}{1+ \frac{1}{\gamma_i+\gamma_1}}\right)\left(1+\frac{1}{\gamma_1+\gamma_N}\right)\right).
\end{equation}
\fi

Thus, the maximum sum rate can be upper bounded by
\begin{equation}\label{upperoptimal}
\Rs(G_O) \leq \frac{1}{2(N-1)}\log_2\left(\prod_{i=1}^{N}{\gamma_i}\times\left({1+ \frac{1}{2\gamma_1}}\right)^N\right).
\end{equation}
Similarly, we can show that for a random pairing $O'$, the corresponding sum rate is lower bounded by
\begin{equation}\label{upperrandom}
\Rs(G_{O'}) \geq \frac{1}{2(N-1)}\log_2\left(\prod_{i=1}^{N}{\gamma_i}\times\left({1+ \frac{1}{2\gamma_N}}\right)^N\right).
\end{equation}
According to (\ref{upperoptimal}) and (\ref{upperrandom}), we find an upper bound for the difference between the sum rate of a random pairing and the optimal pairing as follows
\begin{equation}
\Rs(G_O) - \Rs(G_{O'}) \leq \frac{1}{2}\log_2{\left(\left({\frac{\gamma_N(1+2\gamma_1)}{\gamma_1(1+2\gamma_N)}}\right)^N\right)}
\end{equation}
and as a result
\begin{equation}\label{asympSR}
\lim_{\gamma_1 \to \infty}{\left(\Rs(G_O) - \Rs(G_{O'})\right)} = 0.
\end{equation}
Interestingly, (\ref{asympSR}) shows that for FDF relaying in high SNR regime, the performance of a randomly chosen pairing approaches the performance of the optimal pairing.

\section{Simulation Results} \label{simulations} In this section, we investigate the performance of the optimal pairing in comparison with random pairings. pairing proposed in \cite{Parastoo} is considered as well. We also compare our results with a transmission scheme in which users transmit their data in a time division multiplexing (TDM) fashion. For TDM considered here, uplink is divided into $N$ equal in duration time slots and each user transmits in only one of them. Each uplink is then followed by a downlink in which the relay broadcasts the received message to all users. \textcolor{black}{For TDM, we account for the number of uplink and downlink time slots by considering a pre-log factor of $\frac{1}{2N}$ for the achievable rate of TDM.} Further, the transmit power of each user is scaled such that each user has the same average transmit power as in the optimal pairing. 

We use Monte Carlo simulation to average over common rate and sum rate for the optimal pairing and a randomly selected pairing. For each simulation round, random pairing is selected uniformly at random from all of the feasible client trees. For all users, $P_i$ is set to 1. We assume that transmit power of the relay is proportional to the number of users in the system and set $P_{\mathrm{r}}=N$. Channel gains are assumed to follow Rayleigh distribution and are amplitude samples of a circularly symmetric complex normal random variable $\mathcal{CN}(0,1)$. The number of users is set to $N = 4$.

\if@twocolumn
\begin{figure}[!t]
\centering
\input{CGap.tex}
\includegraphics[width = \columnwidth]{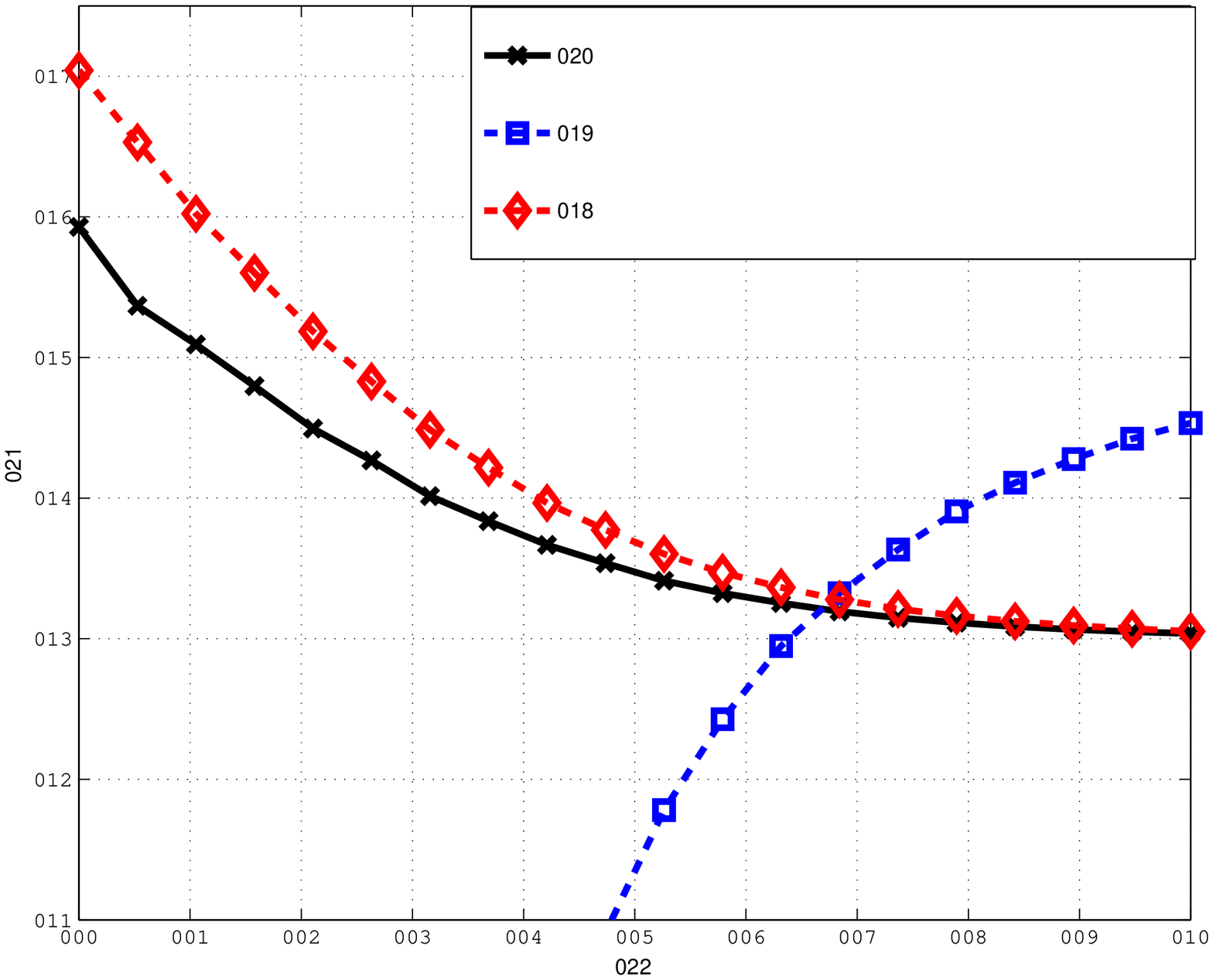}
\caption{Common rate gap wrt TDM transmission, random pairing, and pairing proposed in \cite{Parastoo} for $N = 4$.}
\label{CG}
\end{figure}
\else

\fi
\if@twocolumn
\begin{figure}[!t]
\centering
\input{SGap.tex}
\includegraphics[width = \columnwidth]{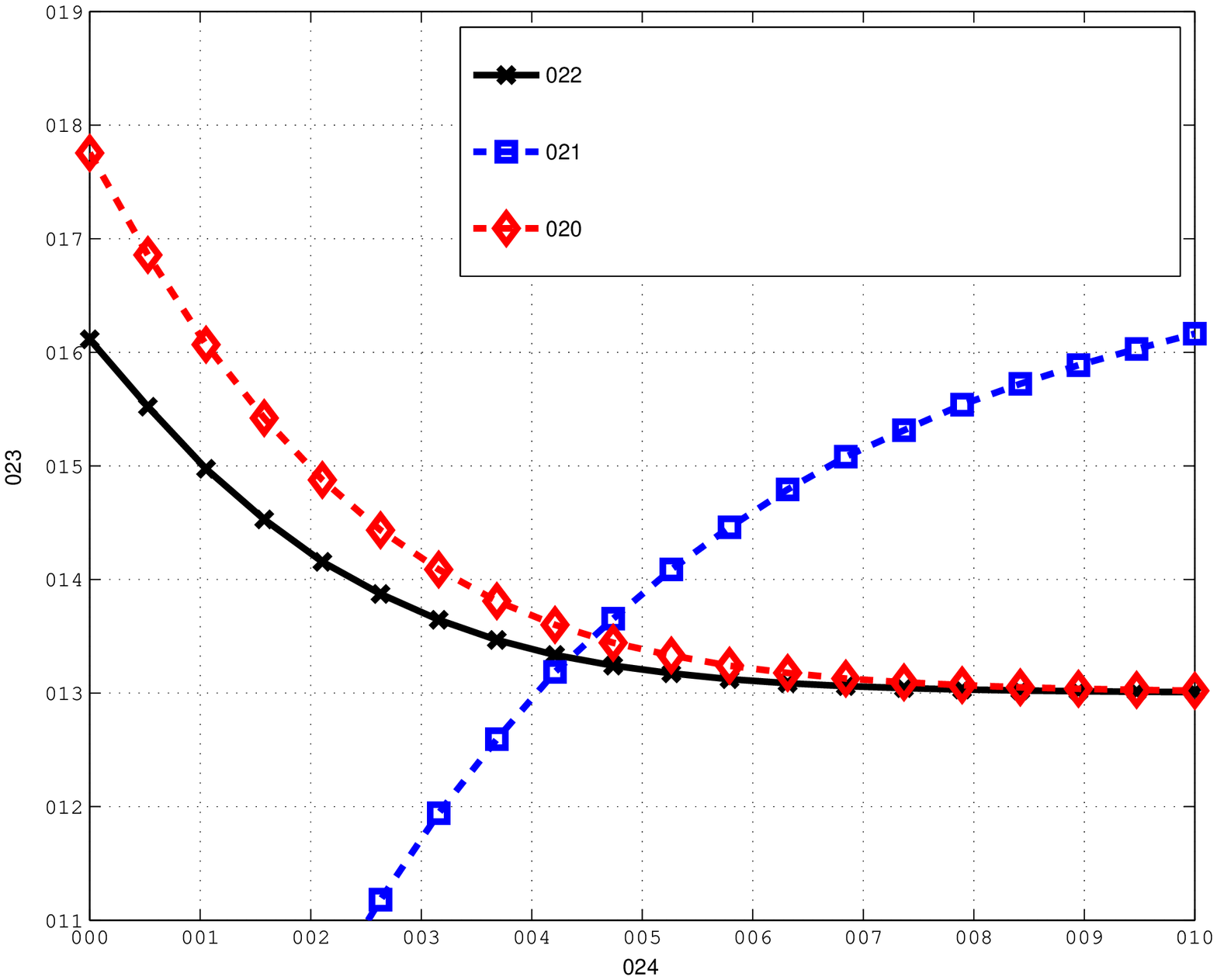}
\caption{Sum rate gap wrt TDM transmission, random pairing, and pairing proposed in \cite{Parastoo} for $N = 4$.}
\label{SG}
\end{figure}
\else
\begin{figure}[ht]
\centering
\begin{minipage}[b]{0.45\linewidth}
\input{CGap2.tex}
\includegraphics[width = \columnwidth]{CGap.eps}
\caption{Common rate gap wrt TDM transmission, random pairing, and pairing proposed in \cite{Parastoo} for $N = 4$.}
\label{CG}
\end{minipage}
\quad
\begin{minipage}[b]{0.45\linewidth}
\input{SGap2.tex}
\includegraphics[width = \columnwidth]{SGap.eps}
\caption{Sum rate gap wrt TDM transmission, random pairing, and pairing proposed in \cite{Parastoo} for $N = 4$.}
\label{SG}
\end{minipage}
\end{figure}
\fi
In order to illustrate the difference between optimal pairing and other pairings, we define the \textit{common rate gap} \cite{Moslem} of the the optimal pairing, $O$, with respect to (wrt) a given pairing, $O'$, as $G_{\mathrm{c}} = 100\times \frac{\Rc(G_O) - \Rc(G_{O'})}{\Rc(G_O)}$ where, by abuse of notation, we denote the average of common rate over all of the simulation rounds by $\Rc (\cdot)$. Similarly, we define the \textit{sum rate gap} as $G_{\mathrm{s}} = 100\times\frac{\Rs(G_O) - \Rs(G_{O'})}{\Rs(G_O)}\ $. 


Fig. \ref{CG} and \ref{SG} illustrate the aforementioned gap and feature the effect of optimal pairing on both common rate and sum rate for FDF relaying. Compared to random pairings, these figures show that the effect of pairing is not significant in higher SNR regimes, as proved earlier. The negative values of the gap function wrt TDM indicates that TDM performs better than pairwise transmission in the low SNR regime.

\begin{figure*}[t]
\centering
\input{FDFbar.tex}
\includegraphics[width=0.9\textwidth]{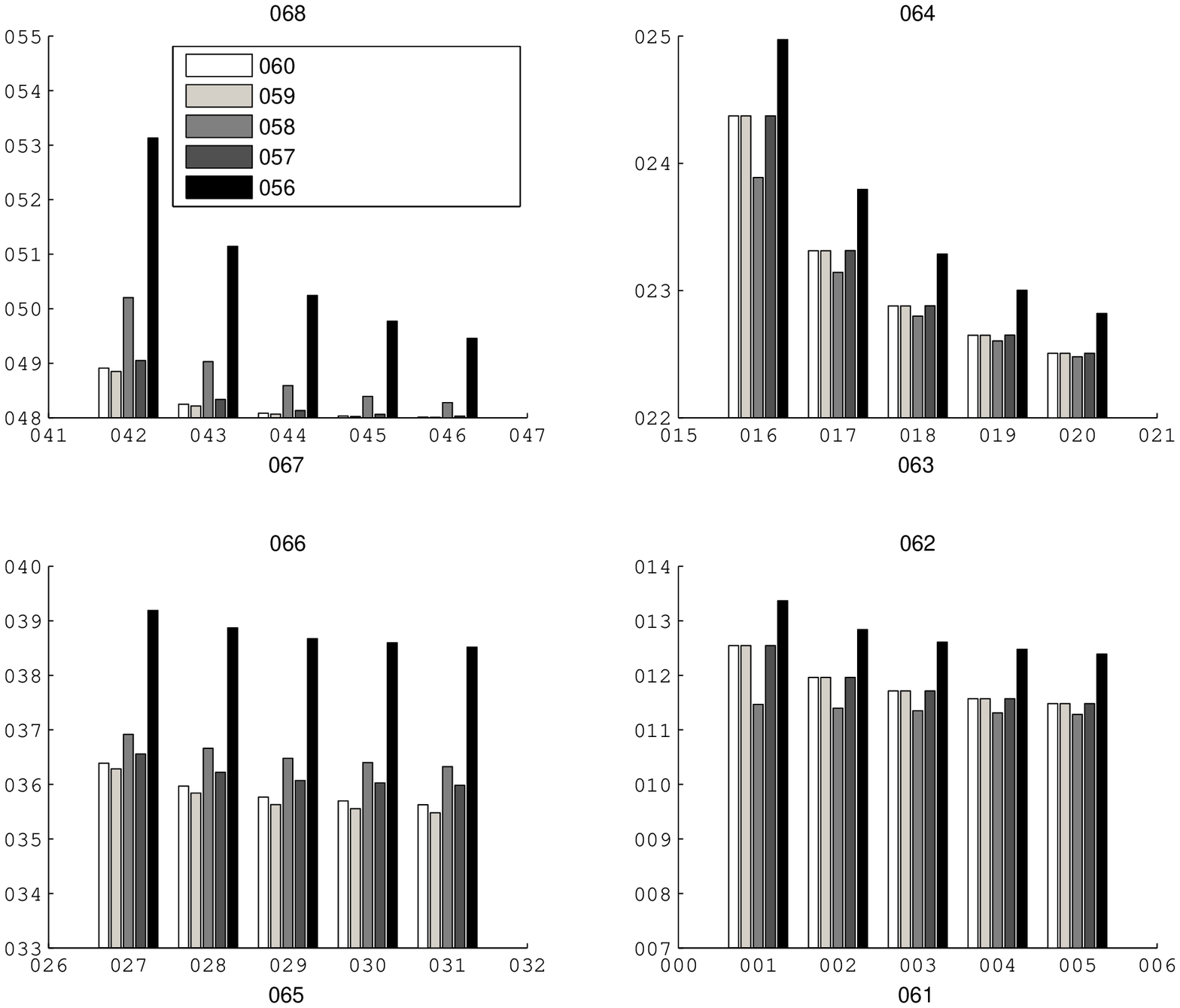}
\caption{Common rate and sum rate of the MWRC with pairwise transmissions and FDF relaying for different number of users, $N$.}
\label{FDFbar}
\end{figure*}

The importance of choosing the optimal pairing may vary when the number of users changes. Fig. \ref{FDFbar} illustrates the performance of optimal pairing, random pairing, and pairing proposed in \cite{Parastoo} in comparison with the cut-set bound (see \cite{Cover}, \cite{Gunduz2009} and Appendix \ref{bounds} for details.) for different numbers of users in a low SNR setting ($1/\sigma^2 = 5$ dB) and a high SNR setting ($1/\sigma^2 = 30$ dB). \textcolor{black}{As seen, for smaller SNRs, the performance improvement of optimal pairing over the random pairing for FDF is more pronounced for larger $N$. This is because as $N$ becomes larger, it is more likely to observe users with highly different channel qualities. This in turn signifies the importance of the pairing as it becomes more important to avoid rate-degrading pairs in the network.}

\vspace{1cm}
\section{Conclusion} \label{conclusion}
In this paper, we studied the effect of users' transmission pairing on the common rate and sum rate of the MWRC with pairwise transmissions and FDF relaying. Optimal pairings were found that maximize common rate and (under a mild practical assumption) sum rate in the system. Moreover, we showed that for high SNR regimes, the effect of pairing becomes less important. Our claims were supported and verified by computer simulations.

\appendices
\section{Proof of Theorem \ref{FDFCR}}\label{appFDFCR}
Here, by an optimal tree, we mean a client tree that achieves the maximum $\Rc$ with respect to (\ref{FDF_rate}). There are two statements regarding (\ref{FDF_rate}) which we use to prove the theorem:
\begin{enumerate}
\item The function $f(x) = x\left(1+\frac{1}{x+ \alpha}\right)$, for $\alpha >0$, is an increasing function of $x$.
\item The function $g(x) = \left(1+\frac{1}{\alpha + x}\right)$ is a decreasing function of $x$.
\end{enumerate}
Given a client tree, $G_O(V,E)$, with an FDF MWRC, we have
\begin{equation}\label{CRFDFdef}
\Rc(G_O) = \min_{i,j} \left\{ \frac{1}{2(N-1)}\log_2\left(\gamma_i+ \frac{\gamma_i}{\gamma_i + \gamma_j}\right)  \right\}.
\end{equation}
where $\ \gamma_i \leq \gamma	_j$ and $ v_iv_j \in E$. Using (\ref{CRFDFdef}), we prove the following lemma. 

\begin{lemma}\label{FDFCRlemma}
There exists an optimal tree, $G_O(V,E)$, in which $A_1^{G_O} = \{v_2\}$.
\end{lemma}
\begin{IEEEproof}
We adapt $G_{O'}(V,E')$ from $G_{O}$ such that we disconnect all of the neighbors of $v_1$ from $v_1$ and connect them to $v_2$. We also make $v_1$ and $v_2$ neighbors. More precisely,
\begin{equation}
E' = (E - \{v_1v_i|v_i \in A_1^{G_O}\}) \cup \{v_2v_i|v_i \in A_1^{G_O}; i \neq 2\} \cup \{v_1v_2\}
\end{equation}
Because of monotonicity of $f(x)$ and $g(x)$, to verify that $\Rc(G_O) \leq \Rc(G_{O'})$, we just need to show
\begin{equation}\label{mideqn}
\gamma_1\left(1+\frac{1}{\gamma_1 + \gamma_{\mathrm{min}}}\right) \leq \gamma_2\left(1+\frac{1}{\gamma_2 + \gamma1}\right)
\end{equation}
where, $\gamma_{min} = \min \{\gamma_i| v_i \in A_1^{G_O}\}$. After some manipulation, we find that (\ref{mideqn}) is equivalent to
\begin{equation}
0 \leq (\gamma_2-\gamma_1)(\gamma_1+\gamma_{\mathrm{min}})(\gamma_2 + \gamma_1) + \gamma_2\gamma_{\mathrm{min}} - \gamma_1^2
\end{equation}
which, according to the fact that $\gamma_1 \leq \gamma_{\mathrm{min}}$, is true.
\end{IEEEproof}
We prove the theorem by induction. If $N = 2$ the theorem obviously holds. Now, assume that the statement of the theorem holds for every FDF MWRC with $N=k$. We show that it also holds for any FDF MWRC with $N=k+1$. For $N=k+1$, according to Lemma \ref{FDFCRlemma}, there exists an optimal tree $G_O(V,E)$ in which $A_1^{G_O} = \{v_2\}$. From equation (\ref{CRFDFdef}), we also have:
\begin{align}\label{CRGO} \nonumber
\Rc(G_O\!) \!  = \! & \min_{  i,j}  \! \! \left\{ \frac{1}{2(N \!\!-\!\!  1\!)} \!  \log_2 \! \! \left( \! \gamma_i \! +\! \! \frac{\gamma_i}{\gamma_i \! +\!  \gamma_j} \!\!\right)\!\!|  \! 1 \! \! < \! i \! \leq \! \!  j ;  v_iv_j \!\!  \in \! \! E \! \right \}\\ 
&\cup \left\{\frac{1}{2(N-1)}\log_2\left(\gamma_1+ \frac{\gamma_1}{\gamma_1 + \gamma_2}\right) \right\}
\end{align}
If the second term in (\ref{CRGO}) is the limiting term in all of the possible client trees with $A_1^{G_O} = \{v_2\}$, the proposed pairing is optimal. Otherwise, maximizing $\Rc(G_O)$ is equivalent to maximizing
\begin{equation}
\min \left\{\gamma_i\left(1+ \frac{1}{\gamma_i + \gamma_j}\right)|1 < i \leq j ; v_iv_j \in E \right\}.
\end{equation}
It is equivalent to maximizing the $\Rc$ for $G_{O'}(V',E')$, in which $V' = V - \{v_1\}$ and $E' = E - \{v_1v_m| v_m \in A_1^{G_O}\}$. According to the induction hypothesis, it happens when 
\begin{equation}
O' = \{\{v_2v_3\}, \{v_3v_4\}, \dots , \{v_{N-1}v_N\}\}
\end{equation}
and as a reslut
\begin{equation}
O = \{\{v_1v_2\}, \{v_2v_3\}, \dots , \{v_{N-1}v_N\}\}
\end{equation}
\hfill$\blacksquare$

\section{Proof of Theorem \ref{FDFSR}}\label{appFDFSR}
Here, we prove Theorem \ref{FDFSR}. 
To prove the theorem, we first show that there is an optimal tree with $deg(v_N)=1$ (Lemma \ref{FDFSR1}). Then we prove that in the optimal tree each node needs to have only one neighbor among nodes with a lower SNR (Lemma \ref{FDFSR2}). We then show that there exist an optimal tree with $deg(v_N) = deg(v_{N-1}) = 1$ (Lemma \ref{FDFSR4}). In the next step, we prove that in an optimal tree for two nodes of degree one, say $v_i$ and $v_j$, if $v_i$ has a higher SNR than $v_j$ then the neighbor of $v_i$ has a higher SNR than the neighbor of $v_j$ (Lemma \ref{FDFSR5}). Then we prove the theorem by induction (Lemma \ref{FDFSR6}).
\begin{IEEEproof}
We use the following convention for the rest of this proof:
\begin{equation}
d_i \triangleq 2^{2(N-1)R_i}.
\end{equation}
As a result, the bound given by (\ref{FDF_weak}) is equivalent to
\begin{equation}
d_i \leq \gamma_i\left(1 + \frac{1}{\gamma_i+\gamma_j}\right).
\end{equation}
We also define $D_s(G_O) = \max\ \prod_{i=1}^{N}{d_i} = 2^{2(N-1) \Rs(G_O)}$. Assume that $G(V,E)$ is a tree such that $\{v_i, v_j, v_k\} \subseteq V$ and$\{v_iv_j, v_iv_k\} \subseteq E $. We define a $\mathcal{V}$-transform on $G$ in such a way that $\mathcal{V}(G,v_i,v_j,v_k) = G'(V, E')$ and $E' = (E - \{v_iv_k\}) \cup \{v_jv_k\}$. Fig. \ref{Vtran} shows the operation of a $\mathcal{V}$-transform.

\begin{figure}[!t]
\psfrag{U_1}{$v_j$}
\psfrag{U_2}{$v_i$}
\psfrag{U_3}{$v_k$}
\psfrag{...}{$\dots$}
\psfrag{arrow}{$\rightarrow$}
\centering
\includegraphics[scale = 0.3]{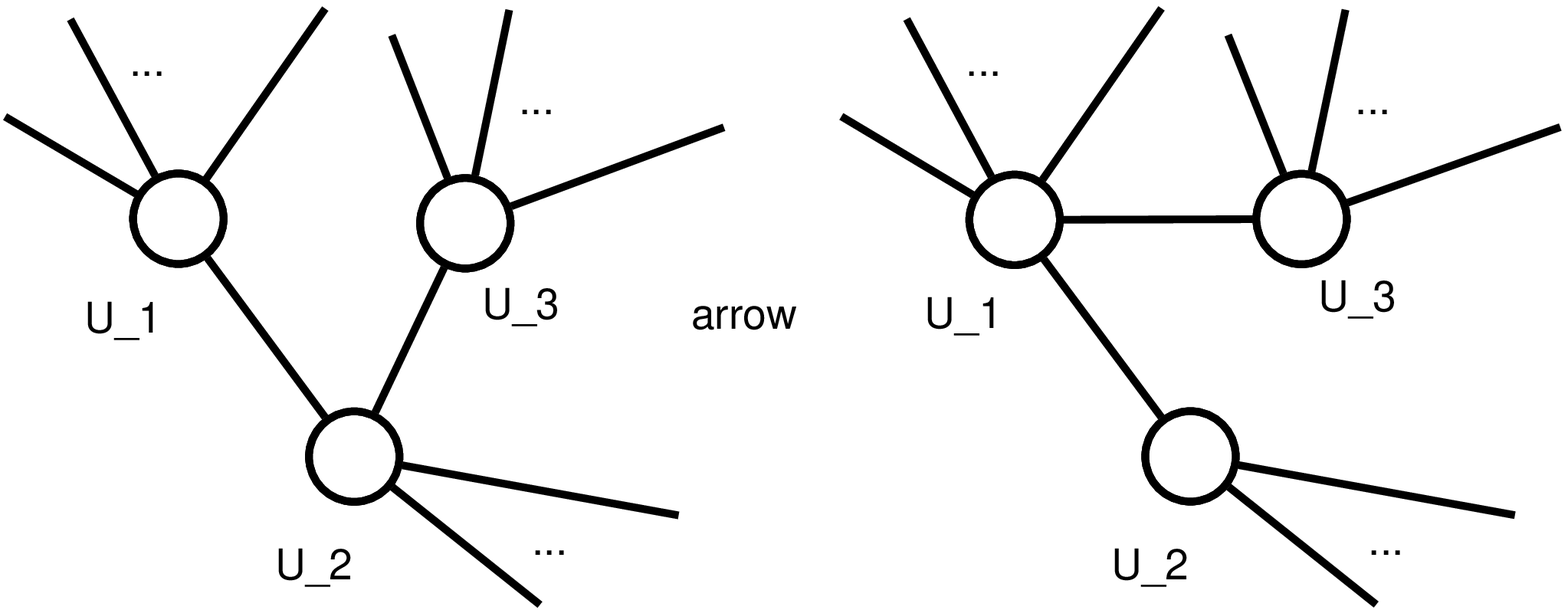}
\caption{Operation of $\mathcal{V}$-transform, $\mathcal{V}(G, v_i,v_j,v_k)$.}
\label{Vtran}
\end{figure}

\begin{lemma}\label{FDFSR1}
There exists an optimal tree in which $deg(v_N) = 1$.
\end{lemma}
\begin{IEEEproof}
Assume $G_O$ is an optimal tree in which $deg(v_N)>1$ and $v_i$ and $v_j$ are two neighbors of $v_N$ and $\gamma_j$ is the minimum SNR value of the neighbors of $V_n$. Consequently, we have $\gamma_i \geq \gamma_j$. It is straightforward to show that by performing a $\mathcal{V}$-transform on $G_O$ and transform it to $G_{O'} = \mathcal{V}(G_O,v_N,v_i,v_j)$,  we have $\frac{D_s(G_{O'})}{D_s(G_O)} \geq 1$:
\if@twocolumn
\begin{equation}\begin{split}
&\frac{D_s(G_{O'})}{D_s(G_O)} \geq \\
&\frac{\left(1+\frac{1}{\gamma_N+\gamma_i}\right)\left(1 + \frac{1}{\gamma_i+h^{G_{O'}}(v_i)}\right)\left(1 + \frac{1}{\gamma_j+h^{G_{O'}}(v_j)}\right)}{\left(1 + \frac{1}{\gamma_i+\gamma_N}\right)\left(1 + \frac{1}{\gamma_i+\gamma_N}\right)\left(1 + \frac{1}{\gamma_j+\gamma_N}\right)} 
\\ &\geq 1.
\end{split}
\end{equation}
\else
\begin{equation}
\frac{D_s(G_{O'})}{D_s(G_O)} \geq
\frac{\left(1+\frac{1}{\gamma_N+\gamma_i}\right)\left(1 + \frac{1}{\gamma_i+h^{G_{O'}}(v_i)}\right)\left(1 + \frac{1}{\gamma_j+h^{G_{O'}}(v_j)}\right)}{\left(1 + \frac{1}{\gamma_i+\gamma_N}\right)\left(1 + \frac{1}{\gamma_i+\gamma_N}\right)\left(1 + \frac{1}{\gamma_j+\gamma_N}\right)} 
\geq 1.
\end{equation}
\fi
Here, $h^{G_{O'}}(v_m)$ is the highest SNR of neighbors of $v_m$ in $G_{O'}$.
This shows that the sum rate of $G_{O'}$ is not less than sum rate of $G_O$.
Note that, after applying this $\mathcal{V}$-transform, we have reduced degree of $v_N$ by one. After applying $deg(v_N) - 2$ more $\mathcal{V}$-transforms, we end up with an optimal tree with $deg(v_N) = 1$. Fig. \ref{Vdeg4} illustrates a hypothetical optimal tree with $deg(v_N) = 4$. It shows how we apply $3$ $\mathcal{V}$-transforms to get an optimal tree with $deg(v_N) = 1$.

\begin{figure}[!h]
\psfrag{U_2}{$v_N$}
\psfrag{U_3}{$v_i$}
\psfrag{...}{$\dots$}
\psfrag{arrow}{$\rightarrow$}
\centering
\includegraphics[scale = 0.3]{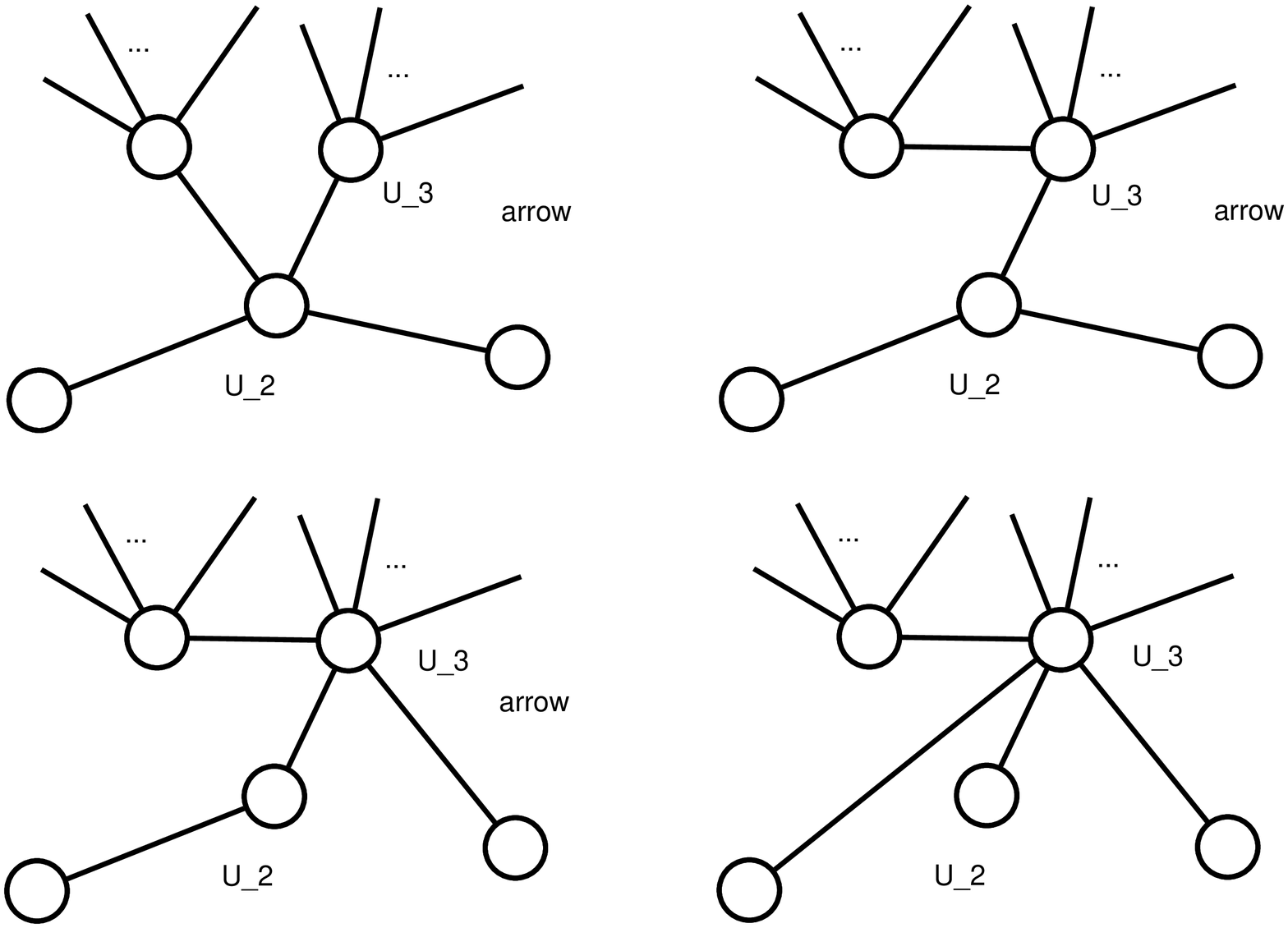}
\caption{Applying 3 $\mathcal{V}$-transform on an optimal tree with $deg(v_N) = 4$.}
\label{Vdeg4}
\end{figure}
\end{IEEEproof}

\begin{lemma}\label{FDFSR2}
There exists an optimal tree, $G_O(V,E)$, such that for any $0<i<N-1$, $deg(v_{N-i}) \leq i+1$. Furthermore, the number of neighbors of $v_{N-i}$ with a lower SNR than $\gamma_{N-i}$ is at most one and consequently, the number of neighbors of $v_{N-i}$ which have higher SNR than $\gamma_{N-i}$ is at least $deg(v_{N-i})-1$.
\end{lemma}
\begin{IEEEproof}
If the number of those neighbors of $v_{N-i}$ that have a lower SNR value than $\gamma_{N-i}$ is $a$, after applying $(a-1)$ $\mathcal{V}$-transforms, we end up with an optimal tree in which $deg(v_{N-i}) \leq i+1$. These $(a-1)$ $\mathcal{V}$-transforms have the form $\mathcal{V}(G,V_{N-i},v_i,v_k)$ and $v_k$ has the highest SNR value among all of the neighbors of $v_{N-i}$.

Now, assume that $deg(v_{N-i}) \leq i+1$ and $v_{N-i}$ has at most one neighbor $v_j$ such that $j < N-i$. Then, we have that the number of neighbors of $v_{N-i}$ that have a higher SNR than $\gamma_{N-i}$ is greater that or equal to $|A_{N-i}^{G_O}|-1 = deg(v_{N-i})-1$.
\end{IEEEproof}

\begin{lemma}\label{FDFSR4}
There exists an optimal tree, $G_O(V,E)$, in which $deg(v_{N}) = deg(v_{N-1}) = 1$. Moreover, if $v_j$ is the only neighbor of $v_{N-1}$ and $v_i$ is the only neighbor of $v_{N}$, then $\gamma_i \geq \gamma_j$.
\end{lemma}
\begin{IEEEproof}
If $deg(v_{N-1}) = 2$, according to Lemma \ref{FDFSR2} and \ref{FDFSR1}, there exists an optimal tree $G_O(V,E)$ in which $deg(v_{N})=1$ and $v_Nv_{N-1} \in E$. Let the other neighbor of $v_{N-1}$ be $v_j$. Then, $G_{O'} = \mathcal{V}(G_O,v_N,v_i,v_j)$ is an optimal tree in which $deg(v_{N-1}) = 1$. So, there always exists an optimal tree $\ G_O$, with $deg(v_{N}) = deg(v_{N-1}) = 1$. Now assume that $deg(v_{N-1})=1$ and the only neighbor of $v_{N-1}$ is $v_{j}$. If $v_{j} = v_{N}$, the graph will be disconnected. Otherwise, if the only neighbor of $v_N$ is $v_i$, we want to prove that $\gamma_i \geq \gamma_j$. We also assume $\gamma_N \neq \gamma_{N-1}$; otherwise, one can rename the nodes in such a way that theorem holds. Assume that $G_{O''}(V,E'')$ is a client tree in which:
\begin{equation}
E'' = (E - \{v_Nv_i, v_{N-1}v_j\}) \cup \{v_Nv_j, v_{N-1}v_i\}.
\end{equation}
We show that $D_s(G_{O''}) \leq D_s(G_{O})$ iff $\gamma_i \geq \gamma_j$:

\begin{equation}
\frac{D_s(G_{O''})}{D_s(G_{O})} = \frac{\left(1+\frac{1}{\gamma_N+\gamma_j}\right)^2\left(1+\frac{1}{\gamma_{N-1}+\gamma_i}\right)^2}{\left(1+\frac{1}{\gamma_N+\gamma_i}\right)^2\left(1+\frac{1}{\gamma_{N-1}+\gamma_j}\right)^2}
\end{equation}
and as a result:
\if@twocolumn
\begin{IEEEeqnarray}{c}\label{proofFDFSR}
\frac{D_s(G_{O''})}{D_s(G_{O})} \leq 1\IEEEnonumber\\
\Leftrightarrow \left(1+\frac{1}{\gamma_N+\gamma_j}\right)\left(1+\frac{1}{\gamma_{N-1}+\gamma_i}\right)\IEEEnonumber\\ 
\leq \left(1+\frac{1}{\gamma_N+\gamma_i}\right)\left(1+\frac{1}{\gamma_{N-1}+\gamma_j}\right)\IEEEnonumber\\
\Leftrightarrow \gamma_N\gamma_j + \gamma_i\gamma_{N-1} \leq \gamma_N\gamma_i + \gamma_{N-1}\gamma_j \IEEEnonumber \\
\Leftrightarrow \gamma_j \leq \gamma_i. \IEEEnonumber
\end{IEEEeqnarray}
\else
\begin{IEEEeqnarray}{c}\label{proofFDFSR}
\frac{D_s(G_{O''})}{D_s(G_{O})} \leq 1\IEEEnonumber\\
\Leftrightarrow \left(1+\frac{1}{\gamma_N+\gamma_j}\right)\left(1+\frac{1}{\gamma_{N-1}+\gamma_i}\right)
\leq \left(1+\frac{1}{\gamma_N+\gamma_i}\right)\left(1+\frac{1}{\gamma_{N-1}+\gamma_j}\right)\IEEEnonumber\\
\Leftrightarrow \gamma_N\gamma_j + \gamma_i\gamma_{N-1} \leq \gamma_N\gamma_i + \gamma_{N-1}\gamma_j \IEEEnonumber \\
\Leftrightarrow \gamma_j \leq \gamma_i. \IEEEnonumber
\end{IEEEeqnarray}
\fi
\end{IEEEproof}
Next lemma, is a generalization of Lemma \ref{FDFSR4} and we prove it in a similar way.
\begin{lemma}\label{FDFSR5}
Assume that $G_O(V,E)$ is an optimal tree in which $deg(v_{N}) = deg(v_{N-1}) = \dots = deg(v_{N-i}) = 1$ and $i < N - 1$. Also, assume that $q < p \leq i$ and $\{v_jv_{N-p}, v_kv_{N-q}\} \in E$. Then $\gamma_j \leq \gamma_k$.
\end{lemma}
\begin{IEEEproof}
It is obvious that $j> N - i$ and $k > N - i$, otherwise the graph is disconnected. Now, if $\gamma_k < \gamma_j$, according to Lemma \ref{FDFSR4}, the graph $G_{O'}(V, E')$ with $E' = (E - \{v_jv_{N-p},v_kv_{N-q}\}) \cup \{v_jv_{N-q}, v_kv_{N-p}\}$ has a greater sum rate which contradicts the fact that $G_O$ is optimal.
\end{IEEEproof}

\begin{lemma}\label{FDFSR6}
Assume $G_O(V,E)$ is an optimal tree and $i$ is the largest integer such that 
\begin{equation}
deg(v_N) = deg(v_{N-1}) = \dots = deg(v_{N-i}) = 1.
\end{equation}
If $ i < N-1$, then there exists an optimal tree $G_{O'}(V,E')$ in which 
\begin{equation}
deg(v_N) = deg(v_{N-1}) = \dots = deg(v_{N-i+1}) = 1.
\end{equation}
\end{lemma}
\begin{IEEEproof}
Assume that $A_{N-i+1}^{G_O} \cap \{v_N, v_{N-1}, \dots, v_{N-i}\} = \{v_{m_1}, v_{m_2}, \dots, v_{m_n}\}$ where $m_1 > m_2 > \dots > m_n$. Define 
\begin{equation}
B = A_{N-i+1}^{G_O} - \{v_N, v_{N-1}, \dots, v_{N-i}\}.
\end{equation}
According to Lemma \ref{FDFSR2}, we assume that $|B| \leq 1$. If $|B| = 0$, $G_O$ is disconnected. Assume $B = \{v_j\}$. Consider $G_{O'}(V,E')$ such that 
\begin{align}\nonumber
E' = &(E -  \{v_{m_1}v_{N-i+1}, v_{m_2}v_{N-i+1}, \dots, v_{m_n}v_{N-i+1}\})\\ &\cup \{v_{m_1}v_j, v_{m_2}v_j, \dots, v_{m_n}v_j\}.
\end{align}
Then, one can conclude that $\frac{D_s(G_O)}{D_s(G_{O'})} \geq 1$ as follows:

\begin{equation}
\frac{D_s(G_O)}{D_s(G_{O'})} \geq \frac{\left(1+\frac{1}{\gamma_{N-i+1}+\gamma_{m_1}}\right)\left(1+\frac{1}{\gamma_{j}+h^{G_{O}(v_j)}}\right)}{\left(1+\frac{1}{\gamma_{N-i+1}+\gamma_{j}}\right)\left(1+\frac{1}{\gamma_{j}+h^{G_{O'}(v_j)}}\right)}
\end{equation}
\begin{equation}
\Rightarrow \frac{D_s(G_O)}{D_s(G_{O'})} \geq \frac{\left(1+\frac{1}{\gamma_{N-i+1}+\gamma_{m_1}}\right)}{\left(1+\frac{1}{\gamma_{N-i+1}+\gamma_{j}}\right)} \geq 1.
\end{equation}

\end{IEEEproof}

According to Lemma \ref{FDFSR6}, there exists an optimal tree with respect to (\ref{FDF_weak}) in which
\begin{equation}
deg(v_N) = deg(v_{N-1}) = \dots = deg(v_{2}) = 1.
\end{equation}
As a result, $O$ is an optimal solution with respect to (\ref{FDF_weak}). The muximum achievable sum rate, $\Rs(G_O)$, could be found directly from (\ref{maxSRFDF}).
\end{IEEEproof}

\section{Cutset Bounds}\label{bounds}
Here, we provide the upper bounds we used In Section \ref{simulations}. For common rate, as it is shown in \cite[Theorem 1]{Moslem}, the common rate for AWGN channels is upper bounded by 
\begin{equation}
\Rc \leq \frac{1}{2(N-1)} \min\left\lbrace \log_2(1+ \sum_{i=1}^{N-1} \gamma_i'), \log_2(1+ \Gamma_{\mathrm{d}} )\right\rbrace.
\end{equation}
This can be found by considering the cut set bound for the cut separating $\{U_2, U_3, \dots , U_N \}$ and the relay in the uplink and also the cut separating $U_1$ and the relay in the downlink. Note that $\gamma_i'$ is the scaled uplink SNR for user $U_i$ such that $U_i$ has the same average transmit power as in the optimal pairing.

For sum rate, we consider the cut separating all users but $U_i$ from the relay for the uplink and also the cut separating $U_i$ from the relay for the downlink. Then we have
\begin{equation}\label{thisone}
\sum_{j \neq i} R_j \leq \min\left\lbrace \frac{1}{2}\log_2(1+ \sum_{j \neq i} \gamma_i'), \frac{1}{2}\log_2(1+ \Gamma_i)\right\rbrace.
\end{equation}
Let $S_i = \min\left\lbrace \frac{1}{2}\log_2(1+ \sum_{j \neq i} \gamma_i'), \frac{1}{2}\log_2(1+ \Gamma_i)\right\rbrace$. Then, by summing over (\ref{thisone}) for all $i = 1,2,\dots, N$ we get
\begin{equation}\label{that}
\sum_{j = 1}^N R_j \leq \frac{1}{N-1}\sum_{j=1}^N S_j.
\end{equation}
We used (\ref{that}) as the upper bound for sum rate in our simulation results.

\makeatother

\section*{Acknowledgment}
The study presented in this paper is supported by TELUS Corporation and Natural Sciences and Engineering Research Council of Canada (NSERC).

\ifCLASSOPTIONcaptionsoff
  \newpage
\fi
\bibliographystyle{IEEEtran}
\bibliography{MyBib}

\end{document}